\RequirePackage{fix-cm}
\documentclass[twocolumn]{svjour3}     

\smartqed  

\usepackage{multicol}

\usepackage{graphicx}
\usepackage{multicol,lipsum}
\usepackage{amsmath}
\usepackage{dcolumn} 
\usepackage{enumitem}
\usepackage{graphicx}
\usepackage{dcolumn}
\usepackage{array,multirow}
\usepackage{bm}
\usepackage{amsmath}
\usepackage{epstopdf}
\usepackage{amsfonts}
\usepackage{amssymb}
\usepackage{mathrsfs}
\usepackage{epsfig}
\usepackage{tabularx}
\usepackage{cite}
\usepackage[dvipsnames]{xcolor}
\RequirePackage[colorlinks,citecolor=blue,urlcolor=blue,linkcolor=blue]{hyperref}
\newcommand{\be}{\begin{equation}}
	\newcommand{\ee}{\end{equation}}
\newcommand{\beq}{\begin{eqnarray}}
	\newcommand{\eeq}{\end{eqnarray}}
	\providecommand{\ed}{\mathrm{d}}

	\usepackage{appendix}

\begin{document}

\title{Probing the parameters of a Schwarzschild black hole surrounded by quintessence and cloud of strings through four standard astrophysical tests}

\author{V.H. Cárdenas \and Mohsen Fathi \and Marco Olivares \and
        J.R. Villanueva 
}

\institute{V.H. Cárdenas \and 
Mohsen Fathi \and J.R. Villanueva 
\at
    Instituto de F\'isica y Astronom\'ia,\\
    Universidad de Valpara\'iso, \\
    Avenida Gran Breta\~na 1111, Valpara\'iso, Chile
\and 
Marco Olivares \at
              Instituto de F\'{\i}sica,\\ Universidad Diego Portales,\\ Avenida Ejercito Libertador, Santiago, Chile         
\and
V.H. Cárdenas
\at
\email{\textcolor{Blue}{\href{mailto:}{victor.cardenas@uv.cl} }}
\and
Mohsen Fathi 
\at
    \email{\textcolor{Blue}{\href{mailto:}{mohsen.fathi@postgrado.uv.cl} }}
\and
Marco Olivares 
\at
    \email{\textcolor{Blue}{\href{mailto:}{marco.olivaresr@mail.udp.cl} }}
\and
J.R. Villanueva
\at
    \email{\textcolor{Blue}{\href{mailto:}{jose.villanueva@uv.cl} }}
}

\date{Received: date / Accepted: date}
\maketitle

\begin{abstract}
In this paper, we concern about applying general relativistic tests on the spacetime produced by a static black hole associated with cloud of strings, in a universe filled with quintessence. The four tests we apply are precession of the perihelion in the planetary orbits, gravitational redshift, deflection of light, and the Shapiro time delay. Through this process, we constrain the spacetime's  parameters in the context of the observational data, which results in about $\sim 10^{-9}$ for the cloud of strings parameter, and $\sim 10^{-20}$ m$^{-1}$ for that of quintessence. The response of the black hole to the gravitational perturbations is also discussed.

{\noindent{\textit{keywords}}: Dark energy, quintessence, cloud of strings, astrophysics}\\

\end{abstract}

\tableofcontents

\section{Introduction and Motivation}\label{sec:intro}

The dark side of the universe has found its way into the physical observations, regarding the flat galactic rotation curves, anti-lensing, and the accelerated expansion of the universe \cite{Rubin1980,Massey2010,Bolejko2013,Riess:1998,Perlmutter:1999,Astier:2012}. This, in fact, has affected the way we look at the astrophysical phenomena. 
Among these, and since the end of the last century, two main observational discoveries have appeared as the keys to obtain a better understating of our universe. First, the confirmation of the highly isotropic black body radiation, of the order $10^{-5}$ of the temperature fluctuations, observed for the cosmic microwave background radiation (CMBR) \cite{Bennett_1994}, and second, the discovery of the accelerated expansion of the universe (in the context of the Friedmann-Lama\^{i}tre-Robertson-Walker (FLRW) metric), using the type Ia supernovae observations \cite{Riess:1998,Perlmutter:1999}. In this context, a concordance model emerges from the observations, which is the so-called Lambda-Cold Dark Matter ($\Lambda$CDM) model.

Despite being simple, this model has been able to give a fairly good description of a wealth amount of the observational data, although its deep theoretical origin is still a mystery, and no clue has been given so far, for the origin and the value of the included cosmological constant. One of the main issues here is the {\it coincidence problem}, or why we live in the exact epoch where the contribution of this constant is of the same order of magnitude as that of matter? In fact, in the extended versions of the model that assume a dynamical source, even no fundamental idea has been put forward to understand this component. 

Nevertheless, there is an approach that has been able to successfully ameliorate the coincidence problem, by replacing the cosmological constant with a quintessence field, in which, the case of an inflaton field during the inflationary epoch, is used as a guide. In order to study the astrophysical phenomena, therefore, it seems logical to consider this model as a conservative approach, since no better explanation exists. Such phenomena may include supernovae, galaxy clusters, or quasars, in addition to which, black hole astrophysics can be named. Black holes, in particular, have appeared among the most interesting astrophysical objects, and the recent imaging of the M87* \cite{Akiyama:2019} has shown that black holes, beside stemming in theoretical concepts, are potentially observable. 

On the other hand, taking into account the cosmological dynamics, the evolution of black holes can also be affected by the dark side of the universe, in which they reside. This process has been discussed extensively in the context of general relativity and alternative theories of gravity (see for example Refs.~\cite{JimenezMadrid:2005,Jamil:2009,Li:2019,Roy:2020}). Geometrically, such calculations would add a dark component to the black hole spacetime under consideration, which is inferred from the cosmological energy-momentum constituents. Such calculations may include the consideration of a dark matter halo \cite{Xu:2018,Das:2021}, or the coupling of the spacetime with a quintessential field \cite{Kiselev:2003,Saadati:2019,AliKhan:2020}.
Furthermore, for the case that the cosmological perfect fluid is regarded as a relativistic dust cloud, consisting of one-dimensional strings (instead of point particles), a specific form of spacetime generalization was done in Ref.~\cite{Stachel:1977}, which associates the black hole to the so-called cloud of strings. 
%
%
This spacetime was generalized further in Ref.~\cite{Letelier:1979} to a gauge-invariant version, and its geodesic structure has been investigated, recently, in Ref.~\cite{Batool:2017}.\\

In this paper, however, we take into account a static black hole spacetime which is associated with both the quintessential field and the cloud of strings. Such black has been derived and discussed in Refs.~\cite{Toledo:2018,Dias:2019,Toledo:2019}, and its geodesic structure regarding the radial orbits has been investigated in Ref.~\cite{Mustafa:2021}. Furthermore, a rotating version of the black was generated in Ref.~\cite{Toledo:2020}, together with discussing its thermodynamics. 
One interesting feature of this black hole spacetime, is that it can include both the effects of dark matter and dark energy, in the sense that the included quintessential component, as well as stemming from the accelerated expansion of the universe, can act as an extra potential granted to the spacetime, to recover the unseen galactic matter. As represented in the next section, such contribution can be found in the Mannheim-Kazanas solution to the fourth order conformal Weyl gravity, that is proposed to recover the flat galactic rotation curves \cite{Mannheim:1989}. The cloud of strings is, however, related to a cosmological model, in which the extended (string-like) objects play role as the sources of gravity, and construct the universe \cite{Letelier:1979}.
%
On the other hand, the respected parameters of the mentioned components are supposed to be appropriately calibrated in the context of standard observations.
According to the fact that such study is missing in the existing literature, in this work, we make the aforementioned black hole to undergo four standard astrophysical tests, in order to be able to constrain the parameters associated with the cloud of strings and quintessence. To elaborate this, in Sect.~\ref{sec:dark}, we give a brief introduction to the spacetime and its components. In Sect.~\ref{sec:tests}, we begin with the precession of perihelion in the planetary orbits as the first test in the solar system. This is followed by the gravitational redshift, gravitational lensing, and the Shapiro time delay. In this section, we use the observational data in the solar system, in order to constrain the spacetime's parameters. As it will be calculated, these values are to small, so that they can appear as perturbations on the Schwarzschild spacetime. To close our discussion, in Sect.~\ref{sec:QNM}, we also present some details about the response of black hole to the gravitational perturbations. We conclude in Sect.~\ref{sec:conclusion}. Throughout this work, we apply a geometrized system of units, in which $G=c=1$.
\section{The black hole solution in the dark background 
}\label{sec:dark}

The static, spherically symmetric black hole solution in the quintessential background, which is surrounded by a cloud of strings, is described by the following metric in the $x^\mu = (t,r,\theta,\phi)$ coordinates:
\begin{eqnarray}\nonumber
&&\ed s^2 =g_{\mu\nu} \ed x^\mu \ed x^\nu \\ \label{eq:metric}
&=& -B(r) \ed t^2 + B^{-1}(r) \ed r^2+r^2 \ed\theta^2+r^2\sin^2\theta \ed\phi^2,
\end{eqnarray}
with the lapse function defined as \cite{Toledo:2018,Dias:2019,Toledo:2019}
\begin{equation}\label{eq:lapse}
B(r) = 1-\alpha - \frac{2M}{r}-\frac{\gamma}{r^{3w_q+1}},
\end{equation}
in which, $\alpha$, $M$, $\gamma$ and $w_q$, represent, respectively, the dimensionless string cloud parameter ($0<\alpha<1$), the black hole mass, the quintessence parameter and the equation of state (EoS) parameter. For a perfect fluid distribution of matter/energy, this latter is defined by $P_q = w_q \rho_q$, with $P_q$ and $\rho_q$ as the quintessential energy pressure and density, and lies within the range $-1<w_q<-{1}/{3}$. This parameter is set to be responsible for the cosmological acceleration and the special case of $w_q=-1$ recovers the cosmological constant. 

To proceed further with our study, we will consider the case of $w_q = -{2}/{3}$ which corresponds to the black hole spacetime with the lapse function
\begin{equation}\label{eq:lapse_1}
B(r) = 1-\alpha - \frac{2M}{r}-\gamma r,
\end{equation}
located in a matter dominated universe \cite{Wei:2008}. 
Note that, the last term resembles the dark matter-related term included in the Mannheim-Kazanas static spherically symmetric solution to the vacuum Bach equations, which is proposed to recover the flat galactic rotation curves \cite{Mannheim:1989}. In this sense, the parameter $\gamma$ can be related to both the dark matter/energy constituents of the spacetime, based on its value (for smaller values, it is mostly related to dark matter).

This spacetime is not asymptotically flat, however, its three-dimensional subspace has an asymptotic deficit of angle \cite{Macias:2002}. Such effect is also intensified by the presence of the cloud of strings. Note that, for this particular choice for the $w_q$, the dimension of $\gamma$ is $\mathrm{m}^{-1}$. 

Defining \cite{Toshmatov:2017,Toledo:2020}
\begin{equation}\label{eq:newParameter}
\rho(r) = M+\frac{\alpha r}{2}+\frac{\gamma r^2}{2},
\end{equation}
for a quintessential energy tensor $T_{\mu\nu} = (\varepsilon,P_r,P_\theta,P_\phi)$ with a constituent of cloud of strings, one can confirm that \cite{Toledo:2020}
\begin{subequations}\label{eq:TmunuComp}
	\begin{align}
	& \varepsilon = \frac{2\rho' }{8\pi }=-P_r,\\
	& P_\theta = P_r-\frac{\rho'' r + 2\rho'}{8\pi r} = P_\phi,
	\end{align}
\end{subequations}
with primes denoting differentiation with respect to the $r$-coordinate, hold in the context of general relativity $G_{\mu\nu} = 8\pi T_{\mu\nu}$, where $G_{\mu\nu}$ is the Einstein tensor. Hence, the solution \eqref{eq:lapse_1} can be regarded as a static black hole spacetime surrounded by a cloud of strings, that is located in a universe filled with quintessential dark energy. Note that, for a comoving time-like observer with a velocity four-vector field $u^\mu = (1,0,0,0)$, the values in Eq. \eqref{eq:TmunuComp} provide
\begin{equation}\label{eq:WEC}
T_{\mu\nu} u^\mu u^\nu = \frac{\alpha +2 \gamma r}{8 \pi  r^2}.
\end{equation}
It is straightforward to verify that for the specific choice of $w_q=-{2}/{3}$, we have $0<\gamma<\frac{(1-\alpha)^2}{8M}\equiv \gamma_c$, and hence, $T_{\mu\nu} u^\mu u^\nu>0$. One can therefore infer that the weak energy condition (WEC) is respected. 
Note that $\gamma_c \rightarrow 0$ for $\alpha \rightarrow 1$, and $\gamma_c = \frac{1}{8M}$ for $\alpha \rightarrow 0$.

This black hole spacetime admits two horizons located at the real roots of the equation $B(r) = 0$, which are
\begin{eqnarray}
&& r_{++} =\frac{1-\alpha}{\gamma}\cos^2\left[\frac{1}{2}\arcsin\left(\frac{2\sqrt{2M\gamma}}{1-\alpha}\right)\right],\label{eq:rg_sB}\\
&& r_+ = \frac{1-\alpha}{\gamma}\sin^2\left[\frac{1}{2}\arcsin\left(\frac{2\sqrt{2M\gamma}}{1-\alpha}\right)\right],\label{eq:rH_sB}
\end{eqnarray}
denoting, respectively, the (quintessential) cosmological, and the event horizons, which will merge to $r_+=r_{++}=r_s=2M$ at the limits $\alpha\rightarrow 0$ and $\gamma\rightarrow 0$. Accordingly, one can re-express the lapse function as $B(r)=\gamma(r-r_+)(r_{++}-r)/r$. Note that, for every specific choice of $\alpha$ within its allowed range, an extremal black hole is obtained for the case of $\gamma = \gamma_c$, with the only horizon located at $r_e = \frac{4 M}{1-\alpha}$, whereas $\gamma > \gamma_c$ corresponds to a naked singularity. \\

In the next section, we continue our discussion by inspecting the astrophysical implications of this black hole spacetime through its  parameters, by means of the observational and experimental data inferred from standard general relativistic tests. These include, the precession of perihelion in the planetary orbits and the deflection of light.

\section{Astrophysical implications
}\label{sec:tests}
In this section, we proceed with comparing the theoretical inferences of doing standard tests on the black hole, with the relevant observational data.  Through this process, one can establish reliable bounds on the  parameters of the spacetime. In what follows, we apply four distinct tests on the black hole, and infer appropriate numerical values of the parameters $\alpha$ and $\gamma$, according to which, the observational and experimental results can be recovered. Note that, since these tests are standard, their explanations can be therefore found in any textbook on general relativity. Hence, we skip the introductory notes and proceed directly to the calculations. We begin with calculating the precession in the perihelion of planetary orbits in the solar system.
%

\subsection{The advance of the perihelion}\label{subsec:perihelion}




An elementary method to study this effect was presented by Cornbleet in Ref.~ \cite{cornbleet93}, which was later applied to other spacetimes in Refs.~ \cite{cov04,Olivares13}. The general idea is to compare the Keplerian elliptic orbits in the Minkowski spacetime (presented in a Lorentzian coordinate system), with those given in the Schwarzschild coordinates. This way, the desired general relativistic corrections are emerged. 
%
%
Let us consider the unperturbed Lorentzian metric 
\begin{equation}
\label{eq:lormet}{\rm d}s^2=-{\rm d}t^2+{\rm d}r^2
+r^2 {\rm d}\theta^2+r^2 \sin^2\theta{\rm d}\phi^2,
\end{equation} 
in the $(t,r,\theta,\phi)$ coordinates, together with metric \eqref{eq:metric}, which we now assume to be in the $(t',r',\theta,\phi)$ coordinates. Accordingly, the relation between $(t,r)$ and $(t',r')$ can be given in the binomial approximations
\begin{subequations}\label{eq:binom}
	\begin{align}
	&  {\rm d}t'=\left(1-\frac{\alpha}{2}-\frac{M}{r}
	-\frac{\gamma}{2}r
	\right){\rm d}t,\\
	&  {\rm d}r'=\left(1+\frac{\alpha}{2}+\frac{M}{r}+
	\frac{\gamma}{2}r
	\right){\rm d}r.
	\end{align}
\end{subequations}
Therefore, in the invariant plane $\theta=\pi/2$, the element of area in the Lorentzian system is $\ed A=\int_0^R r\ed r\ed\phi=\frac{1}{2}R^2\ed\phi$, where $R$ is the areal distance from the planet to the source. This way, the Kepler's second law can be cast as
\begin{equation}
\label{eq:kep1}
\frac{{\rm d}A}{{\rm d}t}=\frac{1}{2}R^2
\frac{{\rm d}\phi}{{\rm d}t}.
\end{equation}
On the other hand, in the Schwarzschild coordinates we have
\begin{eqnarray}
\nonumber{\rm d}A'&=&\int_0^R r{\rm d}r' {\rm d}\phi= \int_0^R \left(r+\frac{\alpha}{2}r+M+
\frac{\gamma}{2}r^2 \right){\rm d}r {\rm d}\phi\\\label{eq:kep2}
&=&\frac{R^2}{2}\left(1+\frac{\alpha}{2}+\frac{2M}{R}+\frac{\gamma}{3}R\right){\rm d}\phi.
\end{eqnarray}
Therefore, by means of the transformations \eqref{eq:binom}, the Kepler's second law is written as
\begin{eqnarray}
\nonumber
&&\frac{{\rm d}A'}{{\rm d}t'}=\frac{1}{2}R^2
\left(1+\frac{\alpha}{2}+\frac{2M}{R}+\frac{\gamma}{3}R\right)
\frac{{\rm d}\phi}{{\rm d}t'}\\\nonumber
&=&\frac{1}{2}R^2
\left(1+\frac{\alpha}{2}+\frac{2M}{R}+\frac{\gamma}{3}R\right) 
\left(1+\frac{\alpha}{2}+\frac{M}{R}+\frac{\gamma}{2}R\right)
\frac{{\rm d}\phi}{{\rm d}t}\\
\label{eq:kep3}
&\simeq&\frac{1}{2}R^2
\left(1+\alpha+\frac{3M}{R}+\frac{4 M \gamma}{3}\right)
\frac{{\rm d}\phi}{{\rm d}t}.
\end{eqnarray} 
In fact, since the law must be held covariant in all coordinate systems, one can infer from Eqs.~\eqref{eq:kep1} and \eqref{eq:kep3}, that $\ed\phi'=(1+\alpha+3M/R+4M\gamma/3)\ed\phi$. Accordingly, for an angular increment $\Delta\phi'$, one gets
%
\begin{equation}
\label{eq:incang}
\int_0^{\Delta \phi'}{\rm d}\phi'= \int_0^{\Delta \phi=2\pi} \left(1+\alpha+\frac{3M}{R}+\frac{4 M \gamma}{3}\right){\rm d}\phi,
\end{equation}
for a single orbit. Knowing that $R=l/(1+\varepsilon \cos \phi)$, for an ellipse with the eccentricity $\varepsilon$ and the semi-latus rectum $l$, one gets
\begin{eqnarray}\nonumber
\Delta \phi'&=&2\pi \left(1+\alpha +\frac{4 M \gamma}{3}\right) +\frac{3M}{l}\int_0^{2\pi} (1+\varepsilon \cos \phi){\rm d}\phi\\
&=& 2\pi +\Delta \phi_{gr}+\Delta \phi_{cs}+\Delta \phi_{q}, 
\end{eqnarray}
where
\begin{subequations}\label{eq:delgr}
	\begin{align}
	&  \Delta \phi_{M} = \frac{6 \pi M}{l},\\
	\label{delcs}
	&  \Delta \phi_{\alpha} = 2\pi \alpha,\\
	\label{delde}
	&  \Delta \phi_{\gamma} = \frac{8\pi M \gamma}{3},
	\end{align}
\end{subequations}
correspond, respectively, to the corrections due to general relativity, cloud of strings and quintessence.

\begin{figure}
	\begin{center}
	\includegraphics[width=8cm]{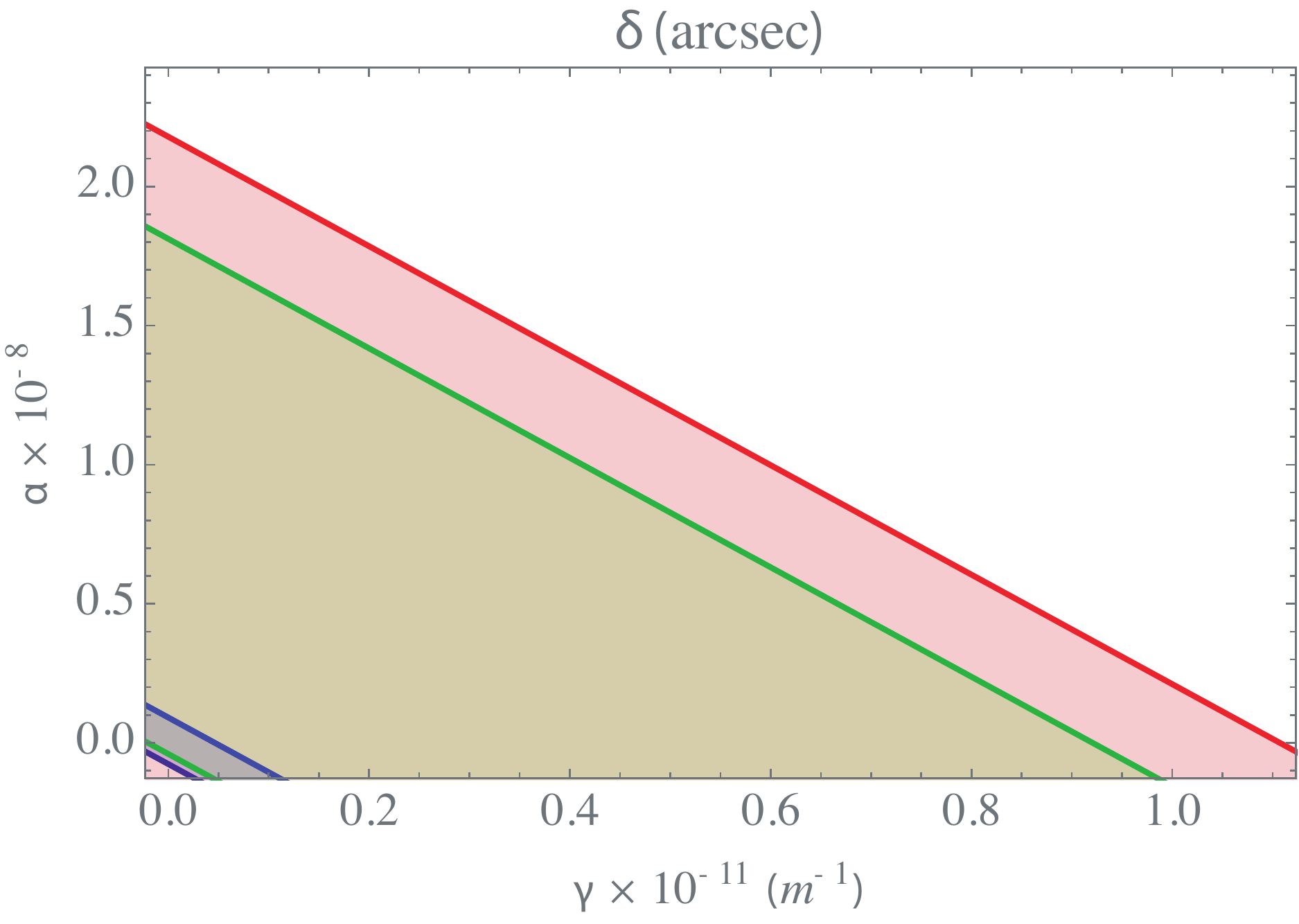}
	\end{center}
	\caption{Constraining the parameters $\alpha$ and $\gamma$, based on the values for the precession in the perihelion of Mercury (blue lines), Venus (green lines), and Earth (red lines) (see Ref.~\cite{cornbleet93} for the respected values).}
	\label{fig:conf.alphagamma.perihelion}
\end{figure}

To test the above relation in the solar system, 
we let $M=M_{\odot}=1476.1$ m, and therefore, the advance of perihelion in arcseconds per century, is obtained as
\begin{equation} \label{eq:adpe}
\delta\equiv \Delta\phi'-2\pi= 573.912 \frac{v}{l}+1.296 v \alpha +2.55072 v \gamma,
\end{equation}
in which, $v$ corresponds to the number of orbits per year, {$l$ is given in $10^9$ m}, $\alpha$ is of order of $10^{-8}$, and $\gamma$ of $10^{-11}$ m$^{-1}$, in accordance with the observed planetary precession in the perihelion in the solar system (see Fig.~\ref{fig:conf.alphagamma.perihelion}).

\subsection{Gravitational redshift}

\begin{figure}
	\begin{center}
	\includegraphics[width=8cm]{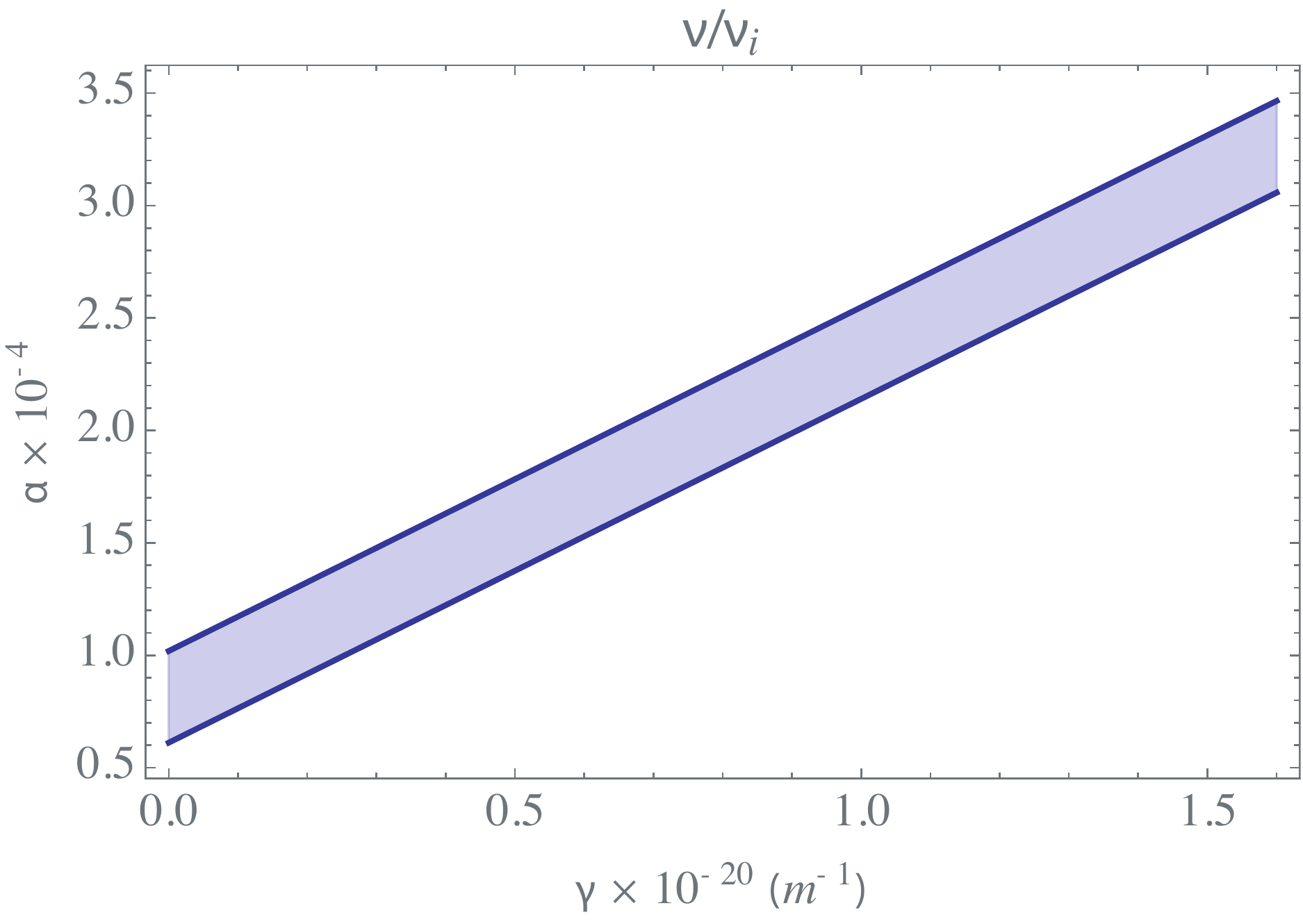}
	\end{center}
	\caption{The confidence range for $\alpha$ and $\gamma$, in accordance with the redshift observed in the GP-A (for the respected values, see Ref.~\cite{Vessot:1980zz}).}
	\label{fig:conf.alphagamma.redshift}
\end{figure}

The famous frequency shift for photons passing a static source, can be inferred from the famous relation \cite{Ryder:2009}
\begin{equation}\label{gredsh}
\frac{\nu}{\nu_i}=\sqrt{\frac{B(r)}{B(r_i)}}.
\end{equation}
which is a result of the existence of a time-like Killing vector associated with the spacetime. Here, $(r_i,\nu_i)$ and $(r,\nu)$ are, respectively, the initial and the observed values of the radial distance to the source and frequency.
%
%
%
%
For the near-earth experiments, however, we have $\alpha\ll 1$ and $\gamma r \ll 2M/r$. One can therefore approximate Eq.~\eqref{gredsh} as
\begin{equation}\label{gredsh2}
\frac{\nu}{\nu_i}\simeq \left(\frac{\nu}{\nu_i}\right)_{\mathrm{gr}}+
\left(\frac{r-r_i}{r_i r}\right) M \alpha-\frac{(r-r_i)}{2} \gamma,
\end{equation}
where 
\begin{equation}
\label{gredsh3}
\left(\frac{\nu}{\nu_i}\right)_{\mathrm{gr}}\equiv 1-\frac{M}{r}+\frac{M}{r_i},
\end{equation}
is the general relativistic value due to the massive source, which has been tested with the hydrogen maser in the Gravity Probe A (GP-A) redshift experiment, with an accuracy of the order of $10^{-14}$ \cite{Vessot:1980zz}. 
Accordingly, the following constraint is obtained:
\begin{equation}
\label{gredsh4} \left|
\left(\frac{r-r_i}{r_i r}\right) M \alpha -\frac{(r-r_i)}{2} \gamma\right|\lesssim 10^{-14}.
\end{equation}
Comparing the initial position $r_i=r_\oplus$ on the Earth of mass $M=M_{\oplus}=4.453\times 10^{-3}$ m, and the observer on a satellite at a height of 15000 km above the Earth, the above relation yields
%
\begin{equation}
\label{gredsh5}|4.877 \alpha-7.5 \gamma| \lesssim 1,
\end{equation}
which constrains $\alpha\sim10^{-4}$ and $\gamma\sim10^{-20} \mathrm{m}^{-1}$ (see Fig.~\ref{fig:conf.alphagamma.redshift}).

\subsection{Deflection of light}

\begin{figure}[t]
	\begin{center}
	\includegraphics[width=8cm]{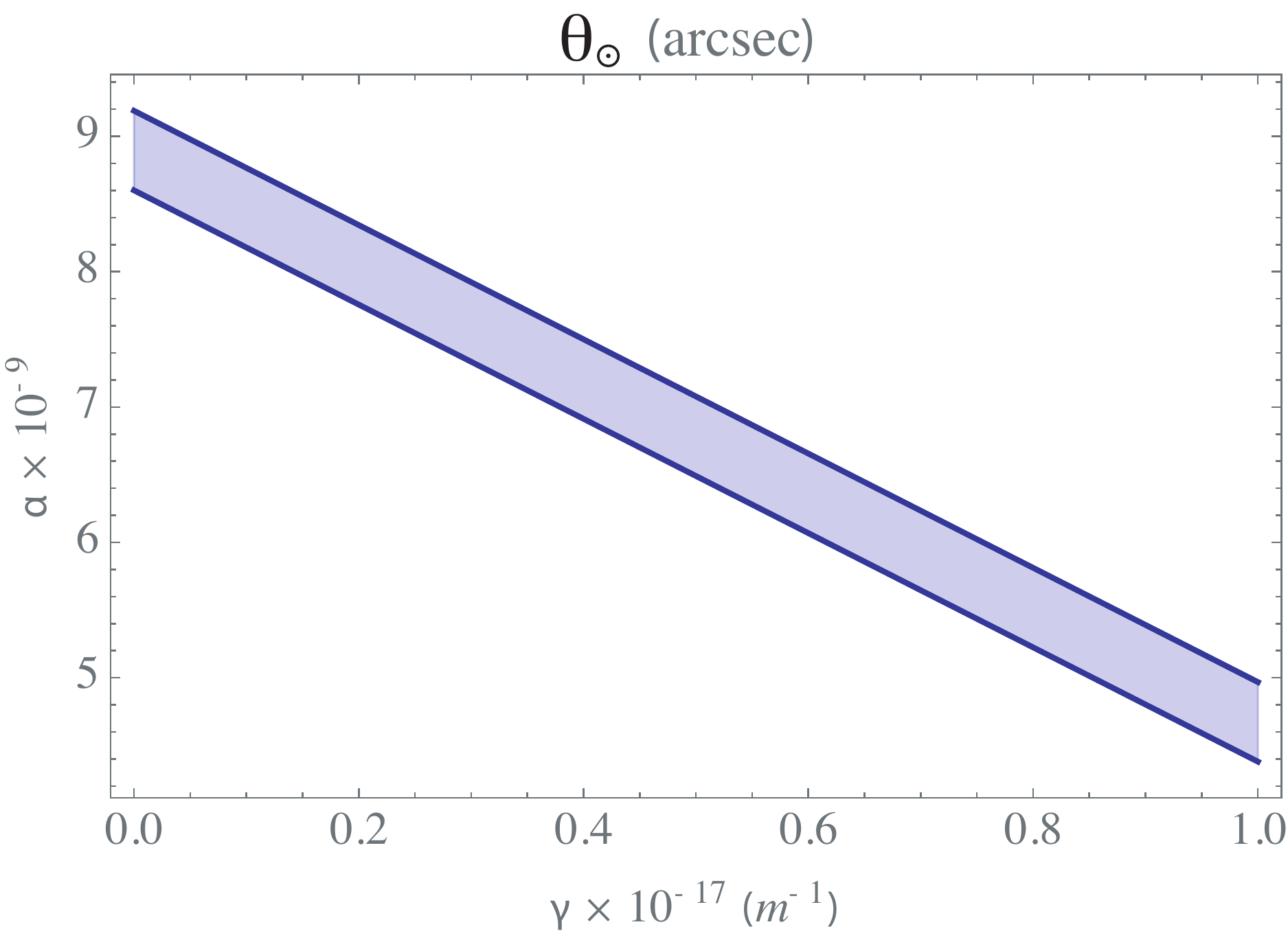}
	\end{center}
	\caption{The constraints on $\alpha$ and $\gamma$ for the deflection angle of the Sun 
	}
	\label{fig:lensigSun}
\end{figure}

The process of light deflection, or the so-called gravitational lensig, can be approached, theoretically, by means of the geodesic equations for the light rays (null geodesics). Indicating $\dot{x}^\mu\equiv{\ed x^\mu}/{\ed s}$, one can get from the line element \eqref{eq:metric} that
\begin{equation}\label{eq:geodesic_0}
\epsilon = -\frac{E^2}{B(r)} \dot{t}^2+\frac{\dot{r}^2}{B(r)} + \frac{L^2}{r^2},
\end{equation}
where $E\equiv B(r)\dot{t}$ and $L\equiv r^2\dot{\phi}$ are the constants of motion, and as in the previous subsections, we have considered the equatorial trajectories corresponding to $\theta = \pi/2$. The parameter $\epsilon$ indicates the nature of the geodesics, in the sense that the null and the time-like trajectories are identified, respectively, by $\epsilon = 0$, and $\epsilon=-1$.
Accordingly, the first order, angular, equation of motion for the light rays (i.e. photons as the test particles) passing the black hole, is given by
%
\begin{equation}
\left(\frac{\dot r}{\dot\phi}\right)^2 = \left(\frac{\ed r}{\ed\phi}\right)^2=\frac{r^{4}}{b^2}- (1-\alpha)r^2+2Mr+\gamma  r^3,
\end{equation}
in which, $b\equiv L/E$ is the impact parameter. Performing the change of variable $r=1/u$, the above equation yields
\begin{equation}\label{ue}
\left(\frac{\ed u}{\ed \phi}\right)^2=\frac{1}{b^2}-(1-\alpha) u^2+2Mu^3+\gamma u,
\end{equation}
that reduces to the standard Schwarzschild equation of light deflection in the limit of $\alpha\rightarrow 0$ and $\gamma 	\rightarrow  0$.
%
Differentiating Eq. \eqref{ue} with respect to $\phi$, gives
\begin{equation}
u^{\prime\prime}+ u =3Mu^2+\alpha u+\frac{\gamma}{2},
\end{equation}
where the primes denote differentiations with respect to $\phi$. Following the procedure established in Ref. \cite{Straumann:2013}, we obtain
\begin{multline}
u=\frac{1}{b}\sin\phi+{ 3M\over 2b^2}+{ \alpha\sqrt{2}\over 2b}+\frac{\gamma}{2}\\
+\left({ M\over 2b^2}+{ \alpha\sqrt{2}\over 12b}\right)\cos(2\phi).
\end{multline}
Note that, $u \rightarrow 0$ results in $\phi\rightarrow\phi_{\infty}$, with
\begin{equation}
-\phi_{\infty}={2M\over b}+{ 7\alpha\sqrt{2}\over 12}+\frac{\gamma b}{2}.
\end{equation}
The deflection angle of the light rays passing the black hole is, therefore, obtained as
\begin{equation}\label{GB1}
\hat{\vartheta} = 2\left |-\phi_{\infty}\right | = {4M\over b}+{ 7\alpha\sqrt{2}\over 6}+\gamma b,
\end{equation}
which recovers the famous form of $\hat{\vartheta}_{\mathrm{Sch}}=4M/b$ for the Schwarzschild black hole in the limits $\alpha	\rightarrow 0$ and $\gamma\rightarrow  0$. This latter, if applied for the Sun as the massive source, provides $\hat{\vartheta}_{\mathrm{Sch}}=4M_{\odot}/R_{\odot}=1.75092 $ arcsec. 
Note that, the observed deflection angle by the Sun has been measured as $\hat\vartheta_\odot = 1.7520$ arcsec for the prograde position, and $\hat\vartheta_\odot = 1.7519$ arcsec for the retrograde one \cite{roy_study_2019}, which produces an error of about 0.0001 arcsec. This error constrains the  parameters as $\alpha\sim 10^{-9}$ and $\gamma\sim10^{-17} \mathrm{m}^{-1}$ (see Fig.~\ref{fig:lensigSun}).

\subsection{Gravitational time delay}
\begin{figure}
	\begin{center}
	\includegraphics[width=8cm]{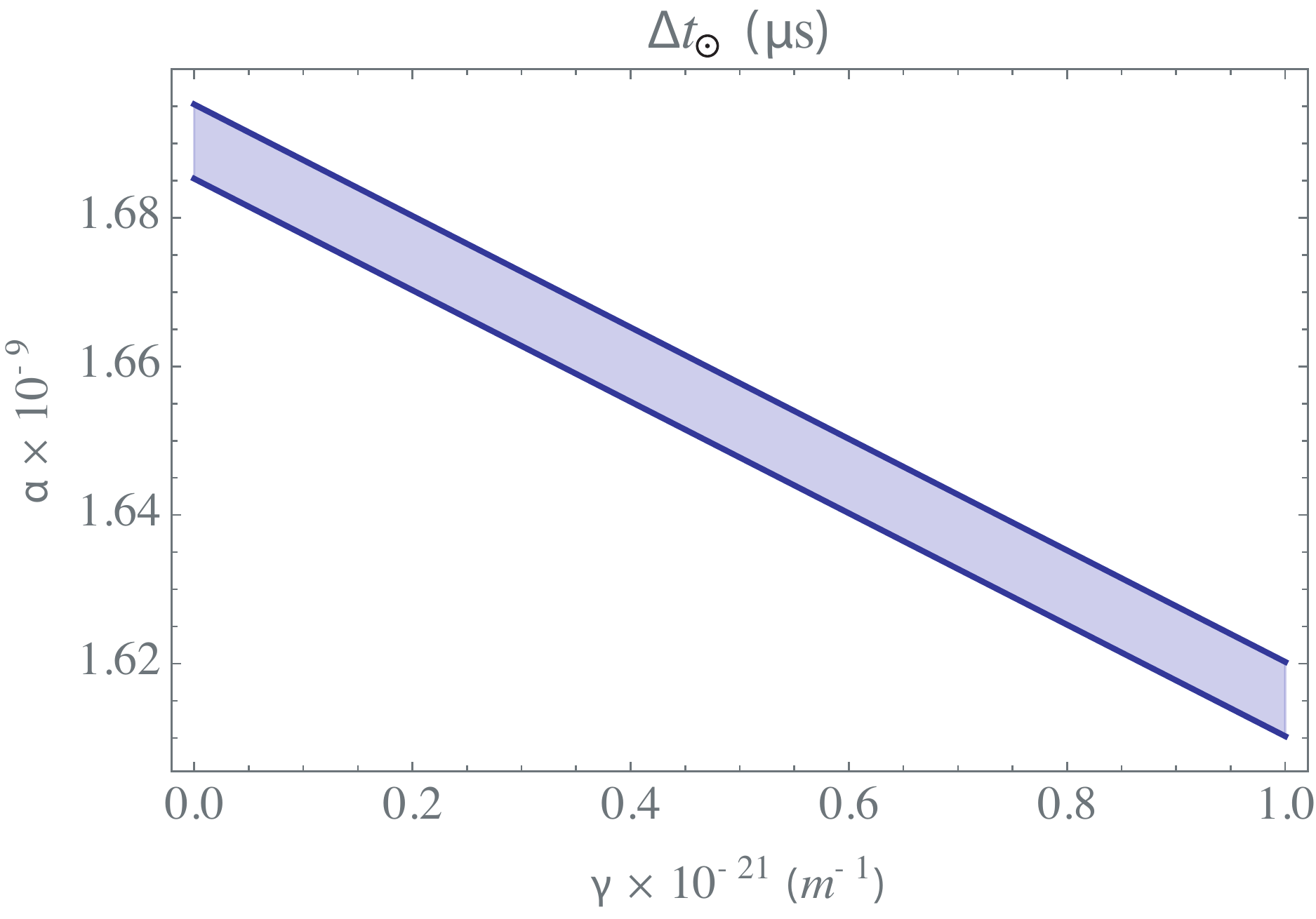}
	\end{center}
	\caption{The constraints of $\alpha$ and $\gamma$ regarding the time delay in the solar system.}
	\label{fig:ShapiroSun_1}
\end{figure}

Claimed as the fourth test of general relativity, the Shapiro time delay has appeared as an interesting effect which is of observational significance. This effect, which refers to the delay in the radar echos of the electromagnetic signals passing massive objects, was proved experimentally by, approximately, the same time of its proposition \cite{shapiro_fourth_1964,shapiro_fourth_1968}. Furthermore, as inferred from recent astrophysical observations, this effect can be seen for two other mass-less energy propagators, namely the neutrinos and the gravitational waves, which act in favor of the existence of dark matter \cite{boran_gw170817_2018}. In this subsection, we proceed with the determination of the resultant Shapiro effect for photons that pass the black hole, by calculating the time difference between the emission and the observation of a light ray, which is sent from the point $P_1=(t_1, r_1)$, travels to $P_2=(t_2,r_2)$, and returns back to $P_1$. Accordingly, we are concerned with the time interval
\begin{equation}
t_{12}=2\, t(r_1,\rho_0)+2\, t(r_2,\rho_0),
\end{equation}
with $\rho_0$ as closest approach to the black hole. Taking into account the definitions given in the previous subsection, we have
\begin{equation}
\dot{r}=\dot{t}\frac{\ed r}{\ed t}=\frac{E}{B(r)}\frac{\ed r}{\ed t},
\end{equation}
from which, one can recast Eq.~\eqref{eq:geodesic_0} as
\begin{equation}\label{ct}
\frac{E}{B(r)}\frac{\ed r}{\ed t}=\sqrt{E^2-\frac{L^2}{r^2}B(r)},
\end{equation}
for mass-less particles. According to the fact that at $r=\rho_0$, the radial velocity of the test particle is vanished, it is straightforward to infer $b^{-2} = B(\rho_0)/\rho_0^2$. This way, the coordinate time is found to vary as
\begin{equation}
t(r,\rho_0)=\int_{\rho_0}^r \frac{\ed r}{B(r)\sqrt{1-\frac{\rho_0^2}{B(\rho_0)}\frac{B(r)}{r^2}}},
\end{equation}
during its journey from $\rho_0$ to $r$. So, to the first order of corrections we obtain
\begin{multline}
t(r,\rho_0)\approx\sqrt{r^2-\rho_0^2}+t_M(r,\rho_0)\\+t_{\alpha}(r,\rho_0)+t_{\gamma}(r,\rho_0),
\end{multline}
where
\begin{subequations}
	\begin{align}
	& t_M(r,\rho_0)=M\left[ \sqrt{\frac{r-\rho_0}{r+\rho_0}}+2\ln\left(\frac{r+\sqrt{r^2-\rho_0^2}}{\rho_0}\right)\right],\label{eq:tM}\\
	& t_{\alpha}(r,\rho_0)=\alpha\sqrt{r^2-\rho_0^2},\label{eq:talpha}\\
	& t_{\gamma}(r,\rho_0)=\gamma \rho_0^2\left[\sqrt{\frac{r-\rho_0}{r+\rho_0}}
	-\ln\left(\frac{r+\sqrt{r^2-\rho_0^2}}{\rho_0}\right)\right]\nonumber
	\\
	&\quad+{\gamma \over 2}\left[r\,\sqrt{r^2-\rho_0^2}
	+\rho_0^2  \ln\left(\frac{r+\sqrt{r^2-\rho_0^2}}{\rho_0}\right)\right].\label{eq:tgamma}
	\end{align}
\end{subequations}
Defining the time difference $\Delta t := t_{12} - t_{12}^{E}$ as the delay for the journey $P_1\rightarrow P_2\rightarrow P_1$, with $t_{12}^{E} = 2\left(\sqrt{r_1^2-\rho_0^2}+\sqrt{r_2^2-\rho_0^2}\right)$ being the travel time interval between the same points in the Euclidean space, one obtains
\begin{equation}\label{eq:Shapiro_0}
\Delta t =\Delta t_M+\Delta t_{\alpha}+\Delta t_{\gamma},
\end{equation}
in which
\begin{subequations}
	\begin{align}
	& \Delta t_M = 2M\left[ \sqrt{\frac{r_1-\rho_0}{r_1+\rho_0}}+\sqrt{\frac{r_2-\rho_0}{r_2+\rho_0}}+2 \ln\left(\frac{\tilde{\mathfrak{t}}_{12}^{E}}{\rho_0^2}\right)\right],\label{eq:DeltatM}\\
	& \Delta t_{\alpha}= \alpha t_{12}^{E},\label{eq:Deltatalpha}\\
	& \Delta t_{\gamma} = 2\gamma\,\rho_0^2\left[\sqrt{\frac{r_1-\rho_0}{r_1+\rho_0}}
	+\sqrt{\frac{r_2-\rho_0}{r_2+\rho_0}}-\ln\left(\frac{\tilde{\mathfrak{t}}_{12}^{E}}{\rho_0^2}\right)
	\right]\nonumber\\
	&+\gamma \left[ r_1\sqrt{r_1^2-\rho_0^2} +r_2\sqrt{r_2^2-\rho_0^2} +\rho_0^2\, \ln\left(\frac{\tilde{\mathfrak{t}}_{12}^{E}}{\rho_0^2}\right)\right]\label{eq:Deltatgamma},
	\end{align}
\end{subequations}
and $\tilde{\mathfrak{t}}_{12}^{E} = \left(r_1+\sqrt{r_1^2-\rho_0^2}\right)\left(r_2+\sqrt{r_2^2-\rho_0^2}\right)$. The expression in Eq.~\eqref{eq:Shapiro_0} is, therefore, the time delay in the echo of light rays passing the black hole. In order to achieve a sensible value for this delay, let us confine ourselves to the solar system, which demands $\rho_0\ll r_1,r_2$. This way, the above difference is approximated as
\begin{multline}\label{eq:Shapiro_1}
\Delta t_\odot \approx 4M\left[ 1+ \ln\left(\frac{4r_1r_2}{\rho_0^2}\right)\right]+2\alpha(r_1+r_2)\\+\gamma\left[ r_1^2+r_2^2 -\rho_0^2 \ln\left(\frac{4r_1r_2}{\rho_0^2}\right)\right].
\end{multline}
Hence, by letting $M\rightarrow M_{\odot}$, $\alpha \rightarrow 0$, and $\gamma	\rightarrow 0$, we recover $\Delta t_{\mathrm{Sch}}=4M_{\odot}\left[ 1+ \ln\left(\frac{4r_1r_2}{\rho_0^2}\right)\right]$, as the Schwarzschild limit of the Shapiro time delay in the solar system. 
Considering $r_1$ and $r_2$ to be, respectively, the Earth-Sun and the Sun-Mars distances, and $\rho_0 \approx R_\odot + (5\times 10^6)$ m, as the approximate radial distance from the Sun's center to its corona, one calculates $\Delta t_{\mathrm{Sch}}\approx 246\,\mathrm{\mu s}$. Note that, the measured error in the observed time difference for the round trip during the Viking mission was about $10\,\mathrm{ns}$ \cite{Viking:1979}. This is related to the confidence values $\alpha\sim10^{-9}$ and $\gamma\sim10^{-21}\mathrm{m}^{-1}$ (see Fig.~\ref{fig:ShapiroSun_1}).\\
%
%

So far, we dealt with some standard general relativistic tests for the black hole, and we constrained the values of the metric parameters, regarding the components of the cloud of strings and quintessence. Since the most important tests have been given sufficient attention, we close this section at this point, and continue our discussion with a more specific concept of the astrophysical black holes. 

\section{Black hole's response to gravitational perturbations and the quasi-normal modes}\label{sec:QNM}

The damping oscillations of the field perturbations in the black hole spacetimes, or the black holes' quasi-normal modes (QNMs), have been of interest among astrophysicists, because of their direct relation to the propagation of the gravitational waves. In fact, the late-time wave form of the black hole ringing is typically identified by a QN frequency \cite{vishveshwara_scattering_1970,press_long_1971,goebel_comments_1972}, which has raised in importance ever since the recent detection of the gravitational waves \cite{PhysRevLett.116.061102,PhysRevLett.116.241103}. The QNMs are therefore absorbing a great deal of attention from the scientific community, since they are also applicable in the gravitational wave astronomy (see for example Refs.~\cite{carter_relativistic_1987,kokkotas_quasi-normal_1999,marranghello_detecting_2007,ferrari_quasi-normal_2008,paschalidis_rotating_2017}). In a more general view, the QNMs are responses of the black holes (or stars) to perturbations. For the Schwarzschild black holes surrounded by a cloud of strings, the QNMs have been calculated in Refs. \cite{doi:10.1142/S0218271817501139,PhysRevD.101.104023}. For scalar perturbations, the scalar QNMs for a Reissner-Nordstr\"{o}m black hole associated with quintessence and cloud of strings have given in Ref.~\cite{Toledo:2019}. In this paper, we continue with calculating the QNMs for the Schwarzschild case, however, we take into account the gravitational perturbations, and confine ourselves to the parameter values that have been determined in the previous subsections. 
For the black hole under consideration, the metric can be perturbed as
\begin{equation}\label{eq:metric2}
\mathfrak{g}_{\mu\nu} = g_{\mu\nu} + h_{\mu\nu},
\end{equation}
according to which, the Einstein equation varies as $ \delta G_{\mu\nu} = 8\pi\delta T_{\mu\nu}$. This perturbation problem can be reduced to a single wave equation, by decomposing it into tensorial spherical harmonics, in the following manner \cite{kokkotas_quasi-normal_1999}:
\begin{equation}\label{eq:Harmonic_0}
\chi(x^\mu) = \sum_{\ell,m}\frac{X_{\ell,m}(t,r)}{r}Y_{\ell,m}(\theta,\phi),
\end{equation}
where the function $X_{\ell,m}(t,r)$ is, in fact, a combination of the all ten independent components of $h_{\mu\nu}$. Note that, since the spacetime under consideration is spherically symmetric, one can omit the index $m$ in the spherical harmonics. Accordingly, we consider the Schr\"{o}dinger-like wave equation 
\begin{equation}\label{eq:perturbEq_0}
\frac{\partial^2 X_\ell}{\partial t^2} - \left(\frac{\partial^2}{\partial r_*^2}-V_\ell(r)\right)X_\ell = 0,
\end{equation}
to govern the radial perturbations outside the event horizon, in which
\begin{equation}\label{eq:r*}
r_* = \frac{r_{+} \ln (r-r_{+})-r_{++} \ln (r_{++}-r)}{\gamma  (r_{++}-r_{+})},
\end{equation}
is the corresponding "tortoise" radial coordinates obeying $\ed r_*=\ed r/B(r)$, and $V_\ell(r)$ is the Regge-Wheeler effective potential \cite{Regge:1957}. The above equation admits two kinds of perturbations, each of which, has an appropriate parity of the effective potential:

\begin{itemize}
	\item For the odd-parity (axial) perturbations, that transform as $(-1)^{\ell+1}$ under the parity transformation, we have
	\begin{equation}\label{eq:Vl-odd}
	V_\ell^-(r) = B(r)\left[\frac{\ell(\ell+1)}{r^2}+\frac{\sigma}{r}B'(r)\right],
	\end{equation}
	where $\sigma=0,1$ and $-3$, correspond, respectively, to the electromagnetic, scalar, and gravitational perturbations.

	\item For the even-parity (polar) perturbations, that transform as $(-1)^{\ell}$ under the parity transformation, we have
	\begin{multline}\label{eq:Vl-even}
	V_\ell^+(r) = \frac{2B(r)}{r^3(3M+kr)^2}
\Big[9M^3+9kM^2 r+3k^2M r^2\\
+k^2(k+1)r^3-9Mr(\alpha+\gamma r)\Big],
	\end{multline}
	where $2k=(\ell-1)(\ell+2)$. For the case of $\alpha=\gamma=0$, the above relation reduces to the Zerilli effective potential for the perturbations on Schwarzschild black hole \cite{Zerilli:1970}. 
\end{itemize}
The potentials have a peak near $r=r_+$, and clearly, they both vanish at the horizons. Considering this, and among several methods in the calculation of the QNMs (see Ref.~\cite{kokkotas_quasi-normal_1999} for a review), we apply the Schutz-Will semi-analytic formula \cite{schutz_black_1985}
\begin{multline}\label{eq:SchutzWill_0}
(M \omega_n)^2 = V_\ell(r_0) - i \left(n+\frac{1}{2}\right)\sqrt{-2\frac{\ed^2 V_\ell(r_0)}{\ed r_*^2}} \\
= V_\ell(r_0)-i\left(n+\frac{1}{2}\right)\sqrt{-2B(r_0)\frac{\ed}{\ed r}\left[B(r_0)\frac{\ed V_\ell(r_0)}{\ed r}\right]},
\end{multline}
which originates from the WKB method of solving the wave scattering problem. Here, $\omega_n$ is the complex QNM frequency, and $r_0$ is the aforementioned potential peak at the vicinity of the event horizon.

\begin{figure}[t]
	\begin{center}
	\includegraphics[width=7cm]{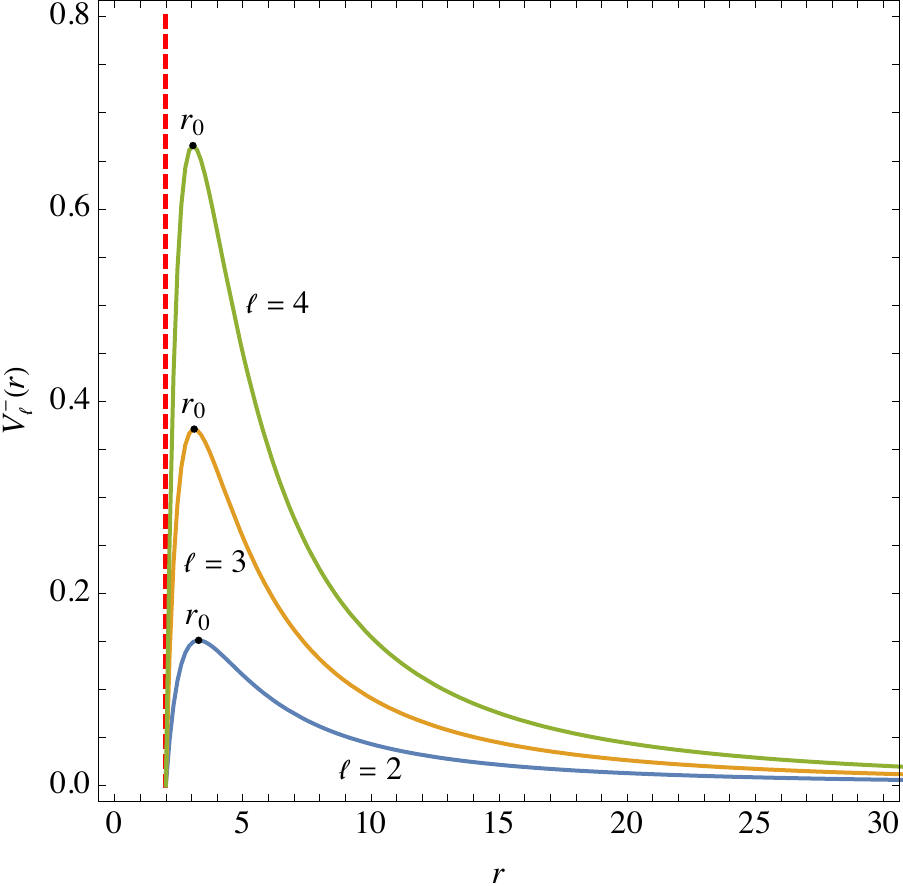}
	\includegraphics[width=7cm]{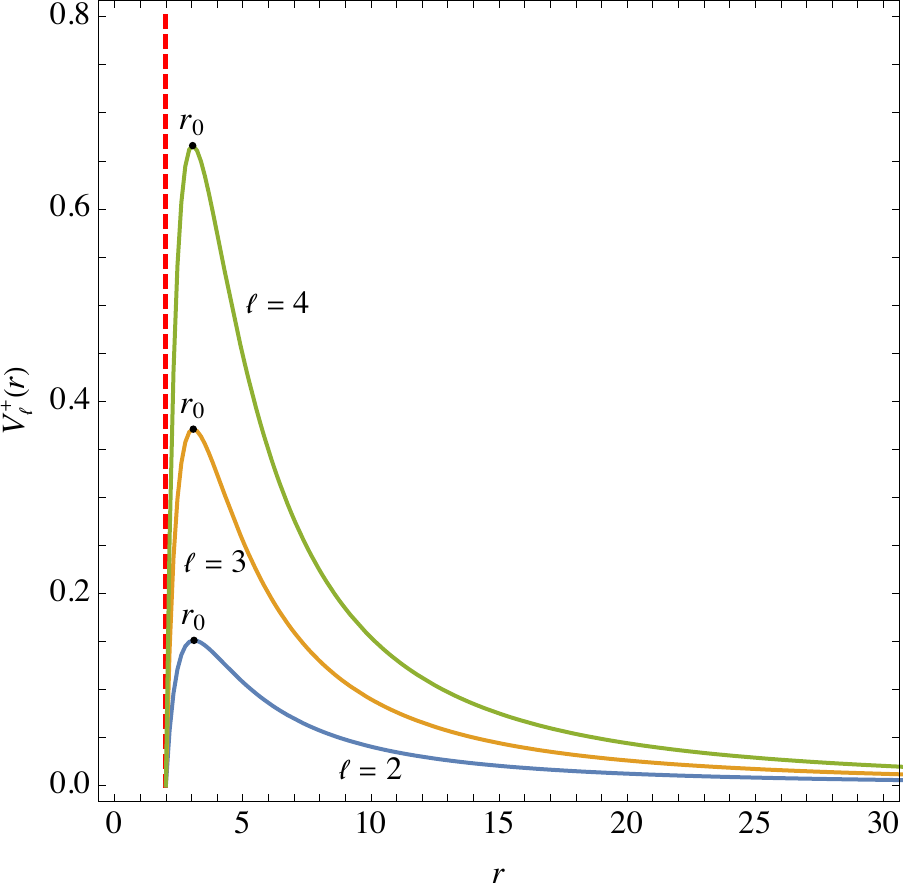}
	\end{center}
	\caption{The Regge-Wheeler effective potentials, plotted for $\alpha=2\times10^{-8}$, $\gamma = 2\times10^{-20}/M$, and the three cases of $\ell = 2,3$ and $4$. The red dashed line indicates the event horizon, and for each of the cases, the potential peak has been indicated by $r_0$.The behavior of the potentials are the same up to 5.54\% of difference. The unit of length along the axes is $M$, and the gravitational perturbations have been taken into account.}
	\label{fig:RW-potentials}
\end{figure}

\begin{figure}
	\begin{center}
	\includegraphics[width=8cm]{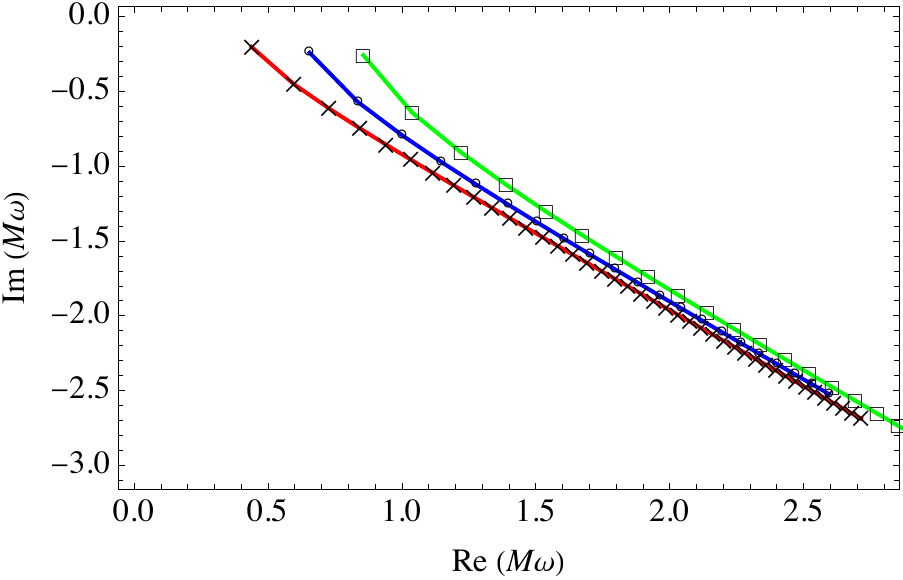}
	\end{center}
	\caption{The spectrum of the QNMs for $\ell=2$ (red), $\ell=3$ (blue), and $\ell=4$ (green).}
	\label{fig:QNMs}
\end{figure}

Let us consider the fundamental mode, that corresponds to $\ell = 2$ and $n=0$. Accordingly, and applying the potential \eqref{eq:Vl-odd} with $\sigma=-3$, we get
\begin{multline}\label{eq:Momegafund}
M\omega_0 = \frac{1}{r_0^2}\Big[
-i r_0 \times\\
\sqrt{{-120 M^2-36 (\alpha -3) M r_0+3 r_0^2 \left[\alpha  (\gamma  r_0+6)+\gamma  r_0-6\right]}}\\
\frac{}{}+12 M^2+6 (\alpha -3) M r_0-3 r_0^2 (\gamma  r_0+2) (\alpha +\gamma  r_0-1)
\Big]^{\frac{1}{2}}.
\end{multline}
The determination of the modes however depends explicitly on the values of $\alpha$ and $\gamma$, which also identify $r_0$ for each of the cases. To elaborate this, we consider Fig.~\ref{fig:RW-potentials}, where we have plotted the potentials given in Eqs. \eqref{eq:Vl-even} and \eqref{eq:Vl-odd}, based on definite values of the metric parameters which have been constrained in the previous subsections in accordance with the observational data, for $\ell=2,3,$ and $4$, and for the case of gravitational perturbations ($\sigma = -3$). 
\begin{table*}[t]
	\begin{minipage}{\textwidth}
		\centering
		\begin{tabular}{c|c|c|c}
			$n$ & $\ell=2$ & $\ell=3$ & $\ell=4$\\
		\hline
			0 & $0.43973\, -0.205123 i$ & $0.65644\, -0.243493 i$ & $0.857432\, -0.260443 i$ \\ 
			1 & $0.597172\, -0.453131 i$ & $0.836703\, -0.573103 i$ & $1.04025\, -0.644016 i$ \\ 
			2 & $0.730026\, -0.617779 i$ & $1.00316\, -0.79668 i$ & $1.22435\, -0.911962 i$ \\ 
			3 & $0.843535\, -0.748508 i$ & $1.14892\, -0.97385 i$ & $1.38999\, -1.1246 i$ \\ 
		\end{tabular}
	\end{minipage}
	\caption{The first four QNMs of the black hole for $\ell = 2,3,$ and $4$, regarding the parameters given in Fig.~\ref{fig:RW-potentials}.}
	\label{table:Momega}
\end{table*}
Based on the small difference revealed from the potentials $V_\ell^\mp(r)$, we take into account the critical distance $r_0$, which is inferred from $V_\ell^-$, reading as $r_0 \approx 3.28 M$. This way, the fundamental mode is calculated as $M\omega_0\approx 0.44-0.21 i$. To infer the corresponding value in kHz, one needs to multiply it by $2\pi(5142 ~\mathrm{Hz}) \times (M_\odot/M)$, which provides the frequency of approximately 1.4 kHz with the damping time 0.66 ms, for a black hole of $M = 10 M_\odot$. The first four QNMs of the black hole have been given in Table \ref{table:Momega}, for $\ell = 2,3$ and $4$. Furthermore, in Fig.~\ref{fig:QNMs}, more modes have been shown in the complex plane, whose number for each value of the harmonic index $\ell$, can be infinite \cite{AIHPA_1993__59_1_3_0,Ferrari:1984}. Also, as it can be seen from the diagrams, the absolute values of the imaginary parts of the frequencies grow rapidly, which implies that the higher modes do not contribute significantly in the emitted gravitational wave signals. This can been seen, as well, in a single mode by growing $\ell$. 

Taking into account the astrophysical constraints we made on the spacetime's  parameters, the above QNMs are the most reliable ones for the black hole, since they relate to the confidence level of the aforementioned parameters. We summarize the general results of this paper in the next section.
%

\section{Summary and the concluding remarks}\label{sec:conclusion}

We studied the astrophysical implications of a Schwarzschild black hole which is associated with cloud of strings and quintessence. This was done by performing standard general relativistic tests in the solar system. 
The corresponding  parameters $\alpha$ and $\gamma$ are supposed to include the effect of extended sources of gravity, as well as dark matter and dark energy.
and the four standard tests could infer the ranges $10^{-9}\leq \alpha \leq 10^{-4}$ and $10^{-21} \leq\gamma M \leq 10^{-11}$. As the smallest values of the  parameters appear inside the confidence range for the experiments related to light propagation in the spacetime, it can be inferred that null trajectories are the most sensitive to changes in these parameters. This, in fact, confirms the pretty well-known observational principle, that the impacts of the possible dark components of the universe, would be first noticeable within the optical and spectroscopic astronomical data. The observational constraints we obtained for this black hole could also pave the way for further studies, in the sense that the physical inferences one obtains can be calibrated within the data reported here. In this paper, also, we calculated the QNMs as the black hole's response to gravitational perturbations, based on particular choices for the  parameters, as the most reliable ones. For higher degrees of $\ell$, each of these modes showed to be of stronger damping, and therefore, of less contribution in the emitted gravitational waves. This feature is in common with other black hole spacetimes, as studied extensively in the literature. For a future work, we aim at studying the thermodynamics of this black hole in the framework of adiabatic processes, so that the observational constraints we presented in this paper can help us having a more realistic vision.

\section*{Acknowledgements}
M. Fathi has been supported by the Agencia Nacional de Investigaci\'{o}n y Desarrollo (ANID) through DOCTORADO Grant No. 2019-21190382, and No. 2021-242210002. J.R. Villanueva was partially supported by the Centro de Astrof\'isica de Valpara\'iso (CAV).

\bibliographystyle{spphys}
\bibliography{biblio_v1.bib}

\end{document}